\pdfoutput=1

\documentclass[11pt]{article}

\usepackage[]{ACL2023}

\usepackage{times}
\usepackage{latexsym}
\usepackage{multicol}
\usepackage{multirow}
\usepackage{adjustbox}
\usepackage{booktabs}
\usepackage{xspace}
\usepackage{amsmath}
\usepackage{amssymb}
\usepackage{tablefootnote}
\usepackage{subfig}
\usepackage{footmisc}
\usepackage{color, colortbl}
\definecolor{LightCyan}{rgb}{0.88,1,1}
\definecolor{LightYellow}{rgb}{1,1,0.88}
\definecolor{LightRed}{rgb}{1,0.88,1}

\newcommand{\se}{StackExchange\xspace}

\newcommand{\eg}{e.g., }
\newcommand{\ie}{i.e., }

\newcommand{\citemissing}[1]{\textcolor{red}{{[citation]}}}

\usepackage{times}
\usepackage{latexsym}

\usepackage[T1]{fontenc}

\usepackage[utf8]{inputenc}

\usepackage{microtype}

\usepackage{inconsolata}

\title{A Survey on Asking Clarification Questions Datasets \\ in Conversational Systems}

\author{Hossein A.~Rahmani$^{\dag}$\thanks{\ \ Equal Contribution} ~ Xi Wang$^{\dag*}$ ~ Yue Feng$^{\dag}$ ~ Qiang Zhang$^{\ddag}$ ~ Emine Yilmaz$^{\dag}$ ~ Aldo Lipani$^{\dag}$ \\ {$^{\dag}$University College London, London, UK} \\ {$^{\ddag}$Zhejiang University, Hangzhou, China} \\ \texttt{{\{hossein.rahmani.22,xi-wang,yue.feng.20,emine.yilmaz,aldo.lipani\}@ucl.ac.uk}} \\ \texttt{{qiang.zhang.cs@zju.edu.cn}}}

\begin{document}
\maketitle
\begin{abstract}
The ability to understand a user's underlying needs is critical for conversational systems, especially with limited input from users in a conversation. Thus, in such a domain, Asking Clarification Questions (ACQs) to reveal users' true intent from their queries or utterances arise as an essential task. However, it is noticeable that a key limitation of the existing ACQs studies is their incomparability, from inconsistent use of data, distinct experimental setups and evaluation strategies. Therefore, in this paper, to assist the development of ACQs techniques, we comprehensively analyse the current ACQs research status, which offers a detailed comparison of publicly available datasets, and discusses the applied evaluation metrics, joined with benchmarks for multiple ACQs-related tasks. In particular, given a thorough analysis of the ACQs task, we discuss a number of corresponding research directions for the investigation of ACQs as well as the development of conversational systems.
\end{abstract}

\section{Introduction}\label{sec:introduction}
Humans often resort to conversations and asking clarification questions to avoid misunderstandings when collaborating with others. Asking Clarification Questions (ACQs) is, therefore, a commonly used mechanism to boost efficiency on human-human as well as human-machine collaborative tasks~\cite{shi2022learning,zou2023asking,shi2023and,feng2023towards}. As an example of human-machine collaboration, conversational systems are developed to not only have a natural conversation with people but also to answer various questions of topics ranging from different domains (e.g.,\ news, movie, and music) in an accurate and efficient manner~\cite{gao2018neural}.
To effectively and efficiently answer various questions, it is essential for many existing conversational systems to capture people's intents.
Only then can conversational systems accurately reply to a series of questions from users~\cite{anand2020conversational,zamani2022conversational}.

Nevertheless, one essential issue is that limited research exists on ACQs and most systems were trained with inconsistent and limited input of data resources. Indeed, in the literature, many studies introduced ACQs to assist conversational systems when applying to different / a mixture of domains (\eg movie~\cite{li2017dialogue} or open domain~\cite{AliannejadiSigir19}). There is also a lack of commonly agreed benchmark datasets for the development of ACQs systems with comparable result analysis. However, on the other hand, in the literature~\cite{AliannejadiSigir19,zamani2020mimics,kumar2020clarq,feng2023towards}, a growing number of studies released publicly available datasets while showing a common interest in the ACQ research direction. This observed contradiction leads to a necessity for a comprehensive overview of the existing datasets as well as the current status of the ACQ research direction. By addressing this concern, many growing ACQs can be better designed, trained and tested with suitable features from properly selected datasets according to comprehensive guidance.


\begin{table*}[!ht]
\centering
\caption{A statistical summary of ACQ datasets for both Conv.~Search and Conv.~QA. The highlighted colours indicate the distinct corpus size of datasets (best viewed in colour). }
\begin{adjustbox}{max width=\textwidth}
\begin{tabular}{lllll}
\toprule
\textbf{Dataset} & \textbf{\# Domains} & \textbf{Scale}  & \textbf{\# Clar.~Q} & \textbf{Link} \\ 
\midrule
\multicolumn{5}{c}{\textbf{Conversational Search}} \\ \hline
\rowcolor{LightRed}
ClariT \cite{feng2023towards}       & - & 108K & 260K &\url{github.com/sweetalyssum/clarit}  \\
\rowcolor{LightCyan}
Qulac \cite{AliannejadiSigir19}       & 198 & 10K & 3K &\url{github.com/aliannejadi/qulac}  \\
\rowcolor{LightCyan}
ClariQ \cite{aliannejadi2021building} & 300 & 2M & 4K & \url{github.com/aliannejadi/ClariQ}   \\
\rowcolor{LightCyan}
TavakoliCQ~\cite{tavakoli2021analyzing} & 3 & 170K & 7K & \url{github.com/Leila-Ta/Clarification_CQA} \\
\rowcolor{LightRed}
MIMICS \cite{zamani2020mimics}        & - & 462K & 586K & \url{github.com/microsoft/MIMICS} \\
\rowcolor{LightYellow}
MANtIS \cite{penha2019introducing}    & 14 & 80K & 435 & \url{guzpenha.github.io/MANtIS/}     \\
\rowcolor{LightCyan}
ClariQ-FKw \cite{sekulic2021towards}  & 230 & 2K & 2K & \url{github.com/isekulic/CQ-generation} \\
\rowcolor{LightYellow}
MSDialog \cite{qu2018analyzing}     & 12 & 35K & 877 &\url{ciir.cs.umass.edu/downloads/msdialog}     \\
\rowcolor{LightCyan}
MIMICS-Dou \cite{tavakoli2022mimics}  & - & 1K & 1K &\url{github.com/Leila-Ta/MIMICS-Duo}     \\
\midrule
\multicolumn{5}{c}{\textbf{Conversational Question Answering}} \\ \hline
\rowcolor{LightRed}
ClarQ \cite{kumar2020clarq}           & 173 & 2M & 2M &\url{github.com/vaibhav4595/ClarQ}  \\
\rowcolor{LightRed}
RaoCQ~\cite{rao2018learning}                & 3 & 77K & 770K &\url{github.com/raosudha89/ranking_clarification_questions} \\
\rowcolor{LightRed}
AmazonCQ~\cite{rao2019answer}                  & 2 & 24K & 179K & \url{github.com/raosudha89/clarification_question_generation_pytorch}  \\
\rowcolor{LightRed}
CLAQUA \cite{xu2019asking}            & 110 & 40K & 40K &  \url{github.com/msra-nlc/MSParS_V2.0} \\
\bottomrule
\end{tabular}
\end{adjustbox}
\label{tbl:datasets} 
\end{table*}

Therefore, in this paper, we offer an overview of the current status of the ACQ research progress. In particular, we aggregate and compare the datasets that have been considered for evaluating recent ACQ techniques from various aspects, such as their dimension, resource, recency and semantic closeness. Afterwards, with the overall discussion of publicly available datasets, we shed light on the model performance while running experiments of corresponding representative techniques on such datasets. Note that, we also release our implementation code for such experiments\footnote{\url{https://github.com/rahmanidashti/ACQSurvey} \label{github}}. Next, we summarised the concluding remarks as well as follow-up suggestions for developing the ACQ techniques.


\paragraph{Our Contributions.} 
The main contributions of this work can be summarized as follows:
\begin{itemize}
    \item We systematically search through $77$ relevant papers, selected as per their recency, reliability and use frequency, in the ACQ domain from top-tier venues.
    \item We compare the ACQ datasets from
    their contributions to the development of ACQ techniques and experimentally show the performance of representative techniques. 
    \item We introduce a visualised semantic encoding strategy to explain dataset suitability when selected for their corresponding experiments.
    \item We analytically outline promising open research directions in the construction of future datasets for ACQs, which sheds light on the development of future research.
\end{itemize}

\section{Conversational Systems}
\label{sec:convsys}

A conversational system functions to assist users while addressing various tasks or acting as a partner in casual conversations~\cite{gao2018neural}. In particular, conversation systems can be classified into four main categories: (1) Conversational Search (Conv.~Search); (2) Conversational Question Answering (Conv.~QA); (3) Task-oriented Dialogues Systems (TDSs); and (4) Social Chatbots~\cite{gao2019neural,anand2020conversational}. In particular, the first two types, \textit{Conv.~Search} and \textit{Conv.~QA}, extend the classic search and QA systems to a conversational nature~\cite{anand2020conversational,zaib2021conversational}. For TDSs and social chatbots, they are more recent research topics and were introduced to build systems for assisting users while addressing a specific task or offering emotional connection and companionship via conversations~\cite{gao2019neural}. However, due to the limited resources that investigate the challenge of asking clarification questions when developing these two systems, this study focuses on Conv.~Search and Conv.~QA systems.
Moreover, ACQs in conversational systems partially focus on three main tasks, namely, Clarification Need Prediction ($T_1$), Asking Clarification Questions ($T_2$), and User Satisfaction with CQs ($T_3$) \cite{zamani2020mimics,tavakoli2022mimics,AliannejadiSigir19}. First, $T_1$ evaluates the necessity of asking clarification questions when users provide their initial queries or requests. Next, with a positive decision, we turn to the action of providing suitable clarification questions (i.e., $T_2$) by following two main routines: generation or selection from a pool of candidate clarification questions. Afterwards, the third task $T_3$ is to evaluate the effectiveness of the corresponding clarification questions while considering user satisfaction levels from multiple aspects (e.g., the usefulness or relevance of clarification questions). An effective ACQ-encoded conversational system requires a joint effort to address the three tasks satisfactorily to enhance users' conversational experience. Therefore, in this survey, we explore the relevant ACQ datasets and discuss their suitability while addressing the above three tasks.

\section{ACQ Datasets}
\label{sec:datasets}

\begin{table*}[!h]
\centering
\caption{A Summary of collection details of ACQ datasets. `-' means that the information is not available. `SE' is \se, `MC' refers to Microsoft Community, and `KB' is Knowledge Base. The detailed information of each dataset, such as the exact source domains, can be accessed in Appendix \ref{sec:datasets_details}.}
\label{tbl:datasets_collection} 
\begin{adjustbox}{max width=\textwidth}
\begin{tabular}{llllll}
\toprule
\textbf{Dataset} & \textbf{Published} & \textbf{Built} & \textbf{Resource} & \textbf{Clar.~Source} \\ 
\midrule
\multicolumn{5}{c}{\textbf{Conversational Search}} \\ \hline
ClariT \cite{feng2023towards}       & 2023 & Aug.~2018 & General queries from task-oriented dialogues & Crowdsourcing\\
Qulac \cite{AliannejadiSigir19}       & 2019 & 2009-2012 & 198 topics from TREC WEB Data & Crowdsourcing\\
ClariQ \cite{aliannejadi2021building} & 2021 & 2009-2014 & 300 topics from TREC WEB Data & Crowdsourcing \\
TavakoliCQ~\cite{tavakoli2021analyzing}          & 2021 & Jul.~2009 to Sep.~2019 & 3 domains of SE & Post and Comment\\
MIMICS \cite{zamani2020mimics}        & 2020 & Sep.~2019 & General queries from Bing users & Machine Generated \\
MANtIS \cite{penha2019introducing}    & 2019 & Mar.~2019 & 14 domains of SE & Post and Comment \\
ClariQ-FKw \cite{sekulic2021towards}  & 2021 & 2009-2014 & TREC WEB Data & Crowdsourcing \\
MSDialog \cite{qu2018analyzing}     & 2018 & Nov.~2005 to Oct.~2017 &  4 domains of MC & Crowdsourcing   \\
MIMICS-Duo \cite{tavakoli2022mimics} & 2022 & Jan.~2022 to Feb.~2022 & General queries from Bing users & HIT on
MTurk, Qualtrics  \\
\midrule
\multicolumn{5}{c}{\textbf{Conversational Question Answering}} \\ \hline
ClarQ \cite{kumar2020clarq}           & 2020 & - & 173 domains of SE & Post and Comment \\
RaoCQ~\cite{rao2018learning}                & 2018 & - & 3 domains of SE & Post and Comment \\
AmazonCQ~\cite{rao2019answer}                  & 2019 & - & A category of Amazon dataset &  Review and Comment \\
CLAQUA \cite{xu2019asking}            & 2019 & - & From an open-domain KB &  Crowdsourcing \\
\bottomrule
\end{tabular}
\end{adjustbox}
\end{table*}

In this section, we describe the main characteristics of the existing and relevant ACQ datasets. Note that we include some additional information, such as the corresponding institution, in Appendix \ref{sec:datasets_details}. A careful dataset selection and aggregation strategy\footnote{We exclude datasets released before 2015 and the ones that are not publicly available.} has been applied to this survey to ensure their recency and accessibility. 

To offer an overview of dataset dimensions, in Table~\ref{tbl:datasets}, we describe the ACQ datasets in statistics, together with links to access the datasets. The statistical information includes the number of the considered domains from the corresponding resource; the size of the whole dataset; the number of clarification questions in each dataset. 
These datasets can be grouped into three sets (large, medium and small, highlighted in pink, cyan and yellow colours) with varied scales of datasets: 1) Large datasets with greater than 10k clarification questions (\ie ClariT, MIMICS, ClarQ, RaoCQ, AmazonCQ, CLAQUA). Note that all the Conv.~QA datasets are classified as large datasets due to the fact that it is more convenient to prepare clarification questions within a QA pair than in a dialogue.
2) Medium datasets with no less than 1K clarification questions (\ie Qulac, ClariQ, TavakoliCQ, ClariQ-FKw, MIMICS-Dou); 3) Small datasets that have no more than 1K instances and only include MANtIS and MSDialog. 
In what follows, we compare datasets for developing conversational search and QA systems, according to their key characteristics.

\subsection{Conversational Search}
Conversational Search (Conv.~Search) refers to information retrieval systems that permit a mixed-initiative interaction with one or more users using a conversational interface~\cite{anand2020conversational}. To develop effective Conv.~Search systems, many previous studies released a number of datasets and made them publicly available. Here, we briefly describe such datasets:

\begin{itemize}
    \item \textbf{ClariT~\cite{feng2023towards}:} The first clarification question dataset for task-oriented information seeking, which asks questions to clarify user requests and user profiles based on task knowledge.

    \item \textbf{Qulac~\cite{AliannejadiSigir19}:} The first clarification question dataset in an open-domain information-seeking conversational search setting with a joint offline evaluation framework. 
    
    \item \textbf{ClariQ~\cite{aliannejadi2020convai3,aliannejadi2021building}:} An extended Qulac with additional crowd-sourced topics, questions and answers in the training corpus as well as synthetic multi-turn conversations.
    
    \item \textbf{TavakoliCQ~\cite{tavakoli2021analyzing,tavakoli2020generating}:} It includes clarification questions collected from the StackExchange QA community and based on three resource categories that have the top number of posts.
    
    \item \textbf{MIMICS~\cite{zamani2020mimics}:} This dataset comprises three sub-datasets that are all sourced from the application of the clarification pane in Microsoft Bing. In particular, they differ in if such a sub-dataset is based on single or multiple clarification panes (\ie MIMICS-Click or ClickExplore) or focusing on real search queries and their corresponding query-clarification pairs (\ie MIMICS-Manual).
    
    \item \textbf{MANtIS~\cite{penha2019introducing}:} A multi-domain (14 domains) conversational information-seeking dataset, sourced from StackExchange, like TavakoliCQ, with joint user intent annotations on the included utterances.
    
    \item \textbf{ClariQ-FKw~\cite{sekulic2021towards}:} This dataset introduces facets (the keywords that disambiguate a query) to the ClariQ, which results in an updated version with a set of query-facet-clarification question triples.
    
    \item \textbf{MSDialog~\cite{qu2018analyzing}:} This dataset was constructed from the dialogues on Microsoft Community\footnote{\url{https://answers.microsoft.com/}} -- a forum that provides technical support for Microsoft products -- and also details user intent types on an utterance level.
    
    \item \textbf{MIMICS-Duo~\cite{tavakoli2022mimics}:} A dataset, stands upon the queries from MIMICS-ClickExplore, that enables both online and offline evaluations for clarification selection and generation approach. 

\end{itemize}

\begin{figure*}[!ht]
    \centering
    \subfloat[\centering tSNE on Conv.~Search Datasets]
    {
        {
            \includegraphics[scale=0.38]{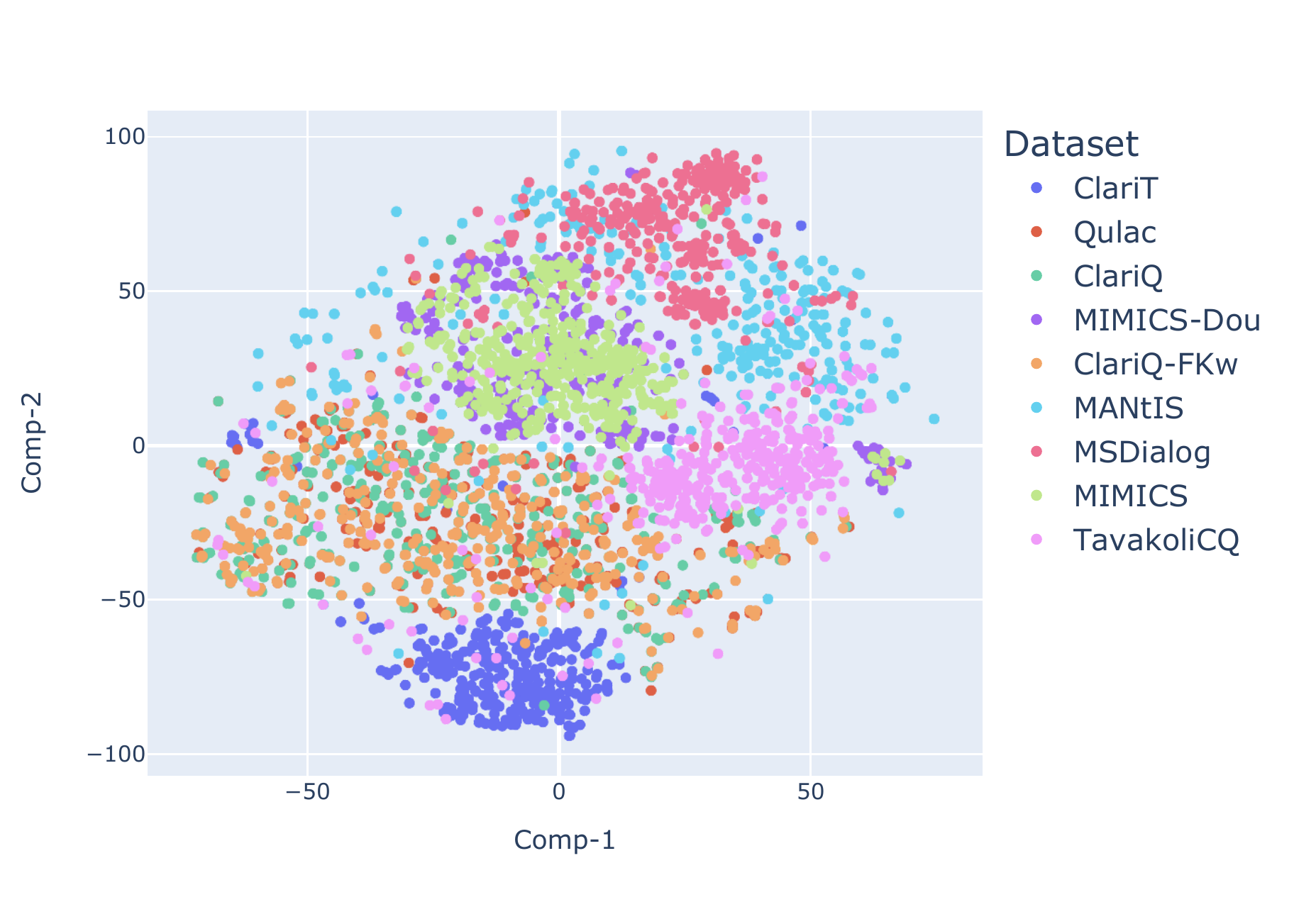}
            \label{fig:tsne_convsearch_datasets}
        }
    }%
    \qquad
    \subfloat[\centering tSNE on Conv.~QA Datasets]
    {
        {
            \includegraphics[scale=0.38]{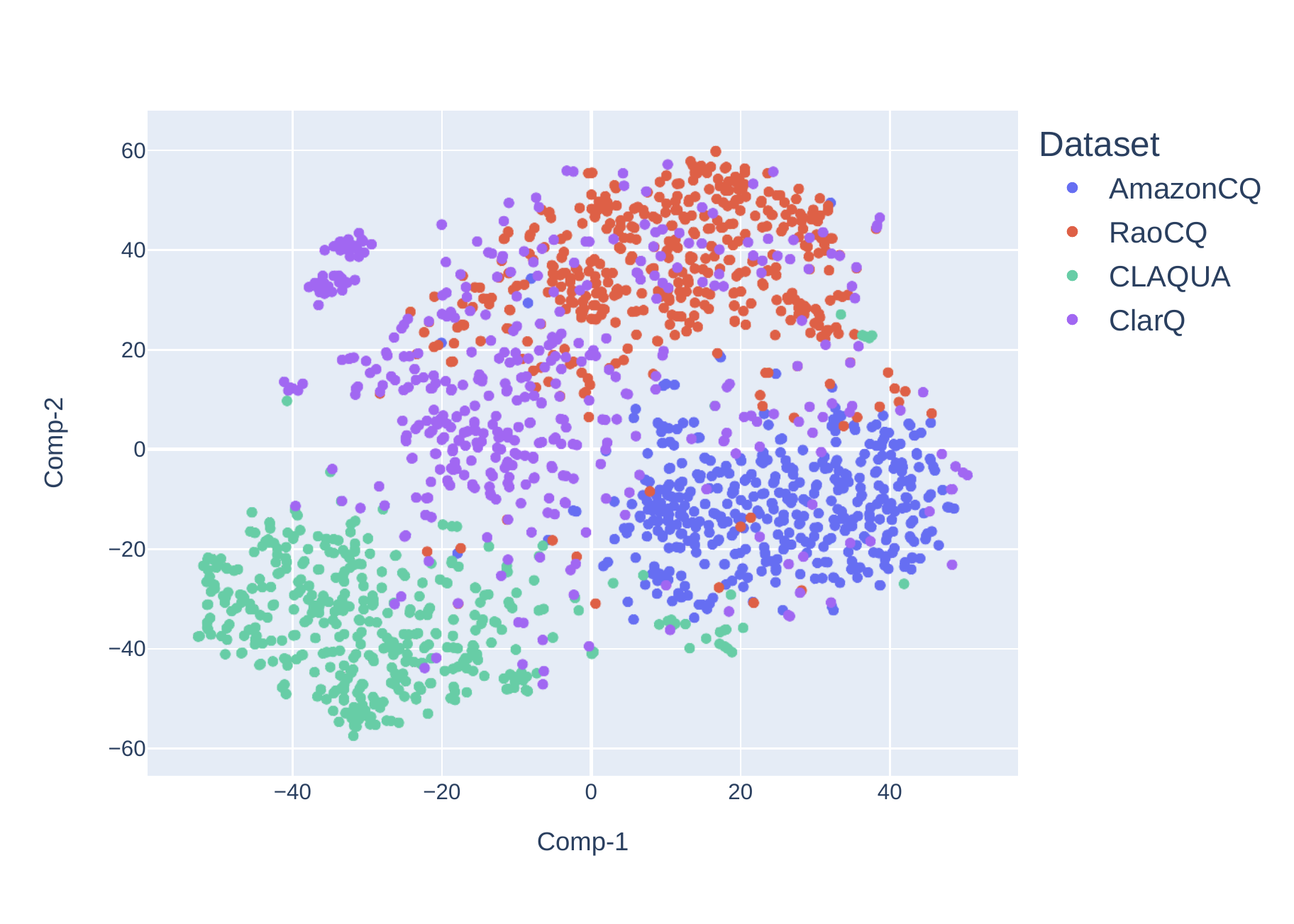}
            \label{fig:tsne_convqa_datasets}
        }
    }%
    \caption{tSNE on ACQ Datasets}%
    \label{fig:tsne}%
\end{figure*}

\subsection{Conversational Question Answering}
The idea behind Conversational Question Answering (Conv.~QA) is to ask the system a question about a provided passage offering a conversational interface \cite{zaib2021conversational}. Conv.~QA has recently received growing attention in the research community while introducing multiple available large-scale datasets. A brief discussion of such datasets are as follows:

\begin{itemize}

    \item \textbf{ClarQ~\cite{kumar2020clarq}:} This dataset is sourced from the post-question pairs in StackExchange and developed with self-supervised approaches within a bootstrapping framework.
    
    \item \textbf{RaoCQ~\cite{rao2018learning}:} Another StackExchange-based dataset with a large volume of post-question-answer triples from three selected domains.
    
    \item \textbf{AmazonCQ~\cite{rao2019answer}:} An Amazon platform-based Clarification QA dataset with questions targeting the missing information of products and answers provided by sellers or other users. In addition, a context is offered that contains both the product title and description.
    
    \item \textbf{CLAQUA~\cite{xu2019asking}:} A clarification-focus dataset that supports the supervised evaluation of text understanding and generation modules, along with a knowledge-based QA system (KBQA).

\end{itemize}

\begin{table}
\centering
\caption{Summary of tasks and evaluation method on ACQs datasets. The tasks can be generation and ranking, which are indicated by `G' and `R', respectively.}
\begin{adjustbox}{width=0.8\columnwidth}
\begin{tabular}{lllll}
\toprule
\multirow{2}{*}{\textbf{Dataset}} & \multicolumn{3}{c}{Task} & \multirow{2}{*}{Eval.~Method} \\ 
\cmidrule{2-4}
 & $T_1$ & $T_2$ & $T_3$ & \\ 
\midrule
\multicolumn{5}{c}{\textbf{Conv.~Search}} \\ \hline
ClariT \citeyearpar{feng2023towards} & $\checkmark$ & G & - & Offline  \\
Qulac \citeyearpar{AliannejadiSigir19} & - & R & - & Offline  \\
ClariQ \citeyearpar{aliannejadi2021building} & $\checkmark$ & R & - & Offline  \\
TavakoliCQ~\citeyearpar{tavakoli2021analyzing} & - & G & - & Offline \\
MIMICS \citeyearpar{zamani2020mimics} & $\checkmark$ & R, G & $\checkmark$ & Offline/Online \\
MANtIS \citeyearpar{penha2019introducing}  & - & R, G & - & Offline  \\
ClariQ-FKw \citeyearpar{sekulic2021towards}  & - & G & - & Offline \\
MSDialog \citeyearpar{qu2018analyzing}  & - & R, G & - & Offline \\
MIMICS-Duo \citeyearpar{tavakoli2022mimics}  & $\checkmark$ & R, G & $\checkmark$ & Offline/Online \\
\midrule
\multicolumn{5}{c}{\textbf{Conv.~QA}} \\ \hline
ClarQ \citeyearpar{kumar2020clarq}  & - & R & - & Offline  \\
RaoCQ~\citeyearpar{rao2018learning} & - & R & - & Offline  \\
AmazonCQ~\citeyearpar{rao2019answer} & - & G & - & Offline \\
CLAQUA \citeyearpar{xu2019asking} & $\checkmark$ & G & - & Offline \\
\bottomrule
\end{tabular}
\end{adjustbox}
\label{tbl:datasets_metrics}
\end{table}

\subsection{Datasets Analysis}
\label{sec:dataset_discussion}
As discussed in Section~\ref{sec:introduction}, a major concern of developing the techniques for asking clarification questions is using suitable datasets to train, validate and test the corresponding approach. In particular, it is essential to be aware of the information on when, how and where a dataset is collected. Such information offers a comprehensive description of datasets for their various characteristics, such as their recency and reliability. Therefore, in Table~\ref{tbl:datasets_collection}, we describe the collection details of each ACQ dataset. In particular, we include the time when the datasets were built as well as the year the corresponding papers were published to indicate the recency of the datasets. In addition, we summarise the source of the data collection, which tells where the datasets came from. Next, we aggregate the main strategies for preparing the clarification questions. At first, due to our data selection strategy, most of the datasets are based on relatively recent information. However, we still observe that some datasets rely on the data collected years ago. For example, the Qulac, ClariQ and ClariQ-FKw datasets consistently use the TREC WEB data but run between 2009 and 2014. The most recent dataset is MIMICS-Duo which was built in 2022, and ClariT is the most recently published dataset in 2023. In particular, all the Conv.~QA datasets are limited, with no time information on when their data was collected, which makes them incomparable based on this measure.
On the other hand, regarding how and where the datasets were collected, the TREC WEB data, StackExchange and Bing are the commonly considered resource for preparing clarification questions in a dataset. Such platforms' search and question-answering nature is the leading cause of such a finding. Afterwards, the crowdsourcing strategy is commonly applied to generate qualified clarification questions. Note that the posts and comments of StackExchange are also widely used to provide clarification questions. According to the provided information, we conclude that the datasets have been collected based on varied strategies, on different periods and use inconsistent resources. However, it is difficult to tell how exactly a dataset is different from others and how to properly select a set of datasets to show the performance of a newly introduced model. Therefore, in this survey, we introduce a visualisation-based approach to assist the selection of datasets for an improved experimental setup.

In Figures~\ref{fig:tsne_convsearch_datasets} and \ref{fig:tsne_convqa_datasets}, we use the t-distributed Stochastic Neighbor Embedding (\ie t-SNE) method to visualize the semantic representation of clarification questions (semantic embeddings) for Conv.~Search and Conv.~QA datasets. As one can see from Figure~\ref{fig:tsne_convsearch_datasets}, Qulac and ClariQ datasets, and MIMICS and MIMICS-Dou datasets highly overlapped with each other. It was expected to be seen as ClariQ and MIMICS-Duo are built on top of Qulac and MIMICS, respectively. This indicates that achieving a high-quality performance of a proposed asking clarification model on both Qulac and ClariQ (or MIMICS and MIMICS-Duo) is not satisfactory as they include clarification questions with close semantic meanings. Figure~\ref{fig:tsne_convsearch_datasets} shows that Conv.~Search datasets form $5$ distinct clusters that can be used to evaluate asking clarification models. For example, the models' generalisability can be evaluated on the ClariT, Qulac, TavakaliCQ, MIMICS, and MSDialog datasets, which locates with few overlapped instances between them. More importantly, comparing Figures~\ref{fig:tsne_convsearch_datasets} and \ref{fig:tsne_convqa_datasets} reveals that clarification questions in Conv.~Search are very focused while the clarification questions in Conv.~QA datasets are more widely distributed. This indicates the high similarities among the Conv.~Search-based data and the resulting necessity of properly selecting those publicly available datasets.
\section{Evaluation Metrics}
\label{sec:metrics}
In this section, we detail the description of the applicable evaluation metrics for the included datasets when evaluating ACQs approaches. In particular, as previously discussed, we discuss such metrics accordingly if they are automatic or human-involved.

\subsection{Automatic Evaluation}
\label{sec:auto_eval}
With a ready dataset, ACQ-based conversational systems can be evaluated using a variety of automatic evaluation metrics. The widely-used metrics can be categorized into two groups based on the strategy of giving clarification questions, \ie ranking or generation. For the ranking route, the commonly used evaluation metrics include (1) MAP \cite{jarvelin2000ir}, (2) Precision \cite{jarvelin2017ir}, (3) Recall \cite{jarvelin2000ir}, (4) F1-score \cite{beitzel2006understanding}, (5) Normalized Discounted Cumulative Gain (nDCG)~\cite{wang2013theoretical}, (6) Mean Reciprocal Rank (MRR)~\cite{voorhees1999trec,radev2002evaluating}, and (7) Mean Square Error (MSE) \cite{beitzel2006understanding}. The main idea behind using these metrics is to evaluate the relevance of the top-ranked clarification questions by the system to reveal the corresponding user intent. On the other hand, some common metrics for the generation route include (8) BLEU \cite{papineni2002bleu}, (9) METEOR \cite{banerjee2005meteor}, (10) ROUGE \cite{lin2004rouge}. 
BLEU and ROUGE were originally developed to evaluate machine translation and text summarization results, respectively. Recently, they have also been applied as evaluation metrics while addressing the ACQ task \cite{sekulic2021towards,zhang2021diverse,shao2022self}. Their scores are both based on the n-gram overlap between generated and reference questions. The difference between BLEU and ROUGE corresponds to the precision and recall metrics. BLEU calculates the ratio of predicted terms in the reference question, while ROUGE scores indicate the ratios of terms from the reference are included in the predicted text. Next, ROUGE-L, a newer version of ROUGE -- focuses on the longest common subsequence -- is recently being used in evaluating ACQ models. However, these above metrics are limited while ignoring human judgements. Therefore the METEOR was introduced to address such a concern by considering the stems, WordNet synonyms, and paraphrases of n-grams.


The main advantage of using automatic evaluation metrics is that they are not expensive for consideration and can be applied easily. However, they are not always aligned with human judgments. Therefore, recent studies also consider human evaluation besides their automatic evaluation to show how the generated or selected CQs impact on the performance of their conversation systems.

\subsection{Human Evaluation}
\label{sec:hu_eval}
In addition to automatic evaluation metrics, human evaluation provides a more accurate and qualitative evaluation of generated or ranked CQs. An essential reason is that automatic evaluation metrics mainly consider n-gram overlaps or ranking of CQs instead of their semantic meaning or other quality-wise aspects. Thus, human annotations are increasingly used to evaluate clarifying questions. The human annotation process consists of scoring generated or selected CQs based on several quality dimensions. Compared to automatic evaluation, human evaluation is naturally more expensive due to the manual annotation effort, but it provides a more accurate picture of the quality of the output. The main aspects that are evaluated using human annotations include (1) \textit{relevance} \cite{aliannejadi2020convai3}, which shows if a CQ is relevant to the user's information need (2) \textit{usefulness} \cite{rosset2020leading} that is related to adequacy and informativeness of a question, (3) \textit{naturalness} \cite{li2019acute} that evaluates a question if it is natural, fluent, and likely generated by a human and (4) \textit{clarification} \cite{aliannejadi2021building} that shows how the user's feedback influences the model's next CQ question. There are also \textit{humanness} \cite{see2019makes}, \textit{engangingness} \cite{li2019acute}, \textit{interestingness} \cite{li2019acute}, \textit{knowledgeable} \cite{li2019acute}, that evaluate a CQ by considering the whole conversation, instead of an individual query-question pair.
However, the ACQ domain lacks a consistent or agreed terminology for the used human evaluation metrics. In addition, some of them could have overlapped focus when evaluating the clarification questions. For example, the \textit{usefulness} can also be evaluated based on the \textit{knowledgeable} of the corresponding clarification question.

\section{Model Performance on ACQ}
\label{sec:performance}

\begin{table}
    \centering
    \caption{Clarification need prediction performance of best representative methods from traditional ML and language models (RandomForest and BERT) on datasets. \textcolor{green}{$\uparrow$} or \textcolor{red}{$\downarrow$} is added to BERT to indicate a consistent performance change on all evaluation metrics. (The results of all methods are added to Table \ref{tbl:clarificationneed_prediction_results} in Appendix \ref{sec:exp_need_prediction_appendix}).}
    \begin{adjustbox}{width=0.8\columnwidth}
    \begin{tabular}{l|ccc}
    \toprule
    \midrule
    \textbf{Model} & \textbf{Precision} & \textbf{Recall} & \textbf{F1}\\
    \midrule
    & \multicolumn{3}{c}{\textbf{ClariQ}}\\
    \midrule
    RandomForest & 0.3540 & \textbf{0.3806} & \textbf{0.3717} \\
    BERT & \textbf{0.3804} & 0.3249 & 0.3344 \\
    \midrule
    & \multicolumn{3}{c}{\textbf{CLAQUA}}\\
    \midrule
    RandomForest & 0.2860 & 0.5000 & 0.3638 \\
    BERT \textcolor{green}{$\uparrow$} & \textbf{0.6349} & \textbf{0.625} & \textbf{0.6255} \\
    \midrule
    \midrule
    \textbf{Model} & \textbf{MAE} & \textbf{MSE} & \textbf{R$^2$}\\
    \midrule
    & \multicolumn{3}{c}{\textbf{MIMICS}}\\
    \midrule
    RandomForest & \textbf{2.4404} & \textbf{7.969} & \textbf{-0.0012} \\
    BERT \textcolor{red}{$\downarrow$} & 2.4562 & 8.1277 & -0.0211 \\
    \midrule
    & \multicolumn{3}{c}{\textbf{MIMICS-Duo}}\\
    \midrule
    RandomForest & \textbf{2.8502} & \textbf{11.206} & \textbf{-0.0079} \\
    BERT \textcolor{red}{$\downarrow$} & 2.8801 & 11.2268 & -0.0098 \\
    \bottomrule
    \end{tabular}
    \end{adjustbox}
    \label{tbl:clarification_need_predication}
\end{table}

In this section, to offer a complete view of the current progress of the ACQ task, we discuss the main observations of the recent ACQ techniques when running on various ACQ datasets. Moreover, for each of the ACQ-related tasks, \ie $T_1$, $T_2$ and $T_3$, we show the performance of many commonly used baselines while running on the applicable datasets for offering some additional concluding remarks.

First, according to our exploration of experimental results of recent ACQ techniques, we observe three main limitations of their inconsistent experimental setups, used baselines and model generalisability. Indeed, many research studies have inconsistent uses of datasets as well as incomparable results with distinct experimental setups. For example, \citet{krasakis2020analysing} and \citet{bi2021asking} both used the Qulac dataset. In \cite{krasakis2020analysing}, they randomly kept 40 topics for testing their performance of a heuristic ranker. However, instead of following~\cite{krasakis2020analysing}, \citet{bi2021asking} used a few-turn-based setup while leveraging the Qulac dataset for asking clarification questions. Next, another common issue is the use of different baselines to show the leading performance of newly introduced techniques. For example, the study in \cite{AliannejadiSigir19} primarily employed ranking-based models, such as RM3, LambdaMART, and RankNet, to evaluate the performance of their question retrieval model. In contrast, the study in \cite{aliannejadi2021building} utilized language models like RoBERTa and ELECTRA to evaluate the performance of their question relevance model. More importantly, many techniques were introduced while tested on a single dataset to show their top performance (e.g., \cite{krasakis2020analysing,sekulic2022exploiting,zhao2022generating}), which lead to a significant generalisability concern. This also indicates the necessity of developing a benchmark while evaluating the ACQ techniques and identifying the exact state-of-the-art. Next, to acquire an overview of model performance while running experiments on the included datasets, we present the experimental results with representative approaches on the three ACQs sub-tasks, \ie $T_1$, $T_2$ and $T_3$ that are discussed in Section~\ref{sec:convsys}. The details of our experiments can be found in Appendix \ref{sec:exp_model_performance}.
\begin{table}[ht]
    \centering
    \caption{Question relevance ranking performance evaluation on representative approaches. `P' and `R' refers to Precision and Recall. \textcolor{green}{$\uparrow$} or \textcolor{red}{$\downarrow$} is added to Doc2Query + BM25 to indicate a consistent performance change to BM25 on all evaluation metrics.} 
    \begin{adjustbox}{width=0.8\columnwidth}
    \begin{tabular}{l|cccc}
    \toprule
    \midrule
    \textbf{Model} &  \textbf{MAP} & \textbf{P@10} & \textbf{R@10} & \textbf{NDCG}\\
    \midrule
    & \multicolumn{4}{c}{\textbf{Qulac}}\\
    \midrule
    BM25 & \textbf{0.6306} & \textbf{0.9196} & 0.1864 & 0.9043\\
    Doc2Query + BM25 & 0.6289	& \textbf{0.9196} & 0.1860 & \textbf{0.9069} \\
     \midrule
     & \multicolumn{4}{c}{\textbf{ClariQ}}\\
     \midrule
     BM25 & 0.6360 & 0.7500 & 0.5742 & 0.7211 \\
     Doc2Query + BM25 \textcolor{green}{$\uparrow$} & \textbf{0.6705} & \textbf{0.7899} & \textbf{0.6006} & \textbf{0.7501} \\
     \midrule
     & \multicolumn{4}{c}{\textbf{TavakoliCQ}}\\
     \midrule
     BM25 & 0.3340 & 0.0463 & 0.4636 & 0.3743 \\
     Doc2Query + BM25 \textcolor{green}{$\uparrow$} & \textbf{0.3781} & \textbf{0.0540} & \textbf{0.5405} & \textbf{0.4260} \\
     \midrule
     & \multicolumn{4}{c}{\textbf{MANtIS}}\\
     \midrule
     BM25 & 0.6502 & 0.0679 & 0.6795 & 0.6582 \\
     Doc2Query + BM25 \textcolor{green}{$\uparrow$} & \textbf{0.7634} & \textbf{0.0830} & \textbf{0.8301} & \textbf{0.7802} \\
     \midrule
     & \multicolumn{4}{c}{\textbf{ClariQ-FKw}}\\
     \midrule
     BM25 & \textbf{0.7127} & 0.5880 & 0.7181 & \textbf{0.7910}\\
     Doc2Query + BM25 & 0.7073 & \textbf{0.5940} & \textbf{0.7244} & 0.7874 \\
     \midrule
     & \multicolumn{4}{c}{\textbf{MSDialog}}\\
     \midrule
     BM25 & \textbf{0.8595} & \textbf{0.0929} & \textbf{0.9293} & \textbf{0.8781}\\
     Doc2Query + BM25 \textcolor{red}{$\downarrow$} & 0.8430 & 0.0908 & 0.9087 & 0.8624 \\
     \midrule
     & \multicolumn{4}{c}{\textbf{ClarQ}}\\
     \midrule
     BM25 & \textbf{0.2011} & \textbf{0.0259} & \textbf{0.2596} & \textbf{0.2200}\\
     Doc2Query + BM25 \textcolor{red}{$\downarrow$} & 0.1977 & 0.0263 & 0.2630 & 0.2168 \\
     \midrule
     & \multicolumn{4}{c}{\textbf{RaoCQ}}\\
     \midrule
     BM25 & \textbf{0.1511} & 0.0236 & 0.2362 & 0.1797 \\
     Doc2Query + BM25 & 0.1509 & \textbf{0.0241} & \textbf{0.2415} & \textbf{0.1811} \\
     \midrule
     & \multicolumn{4}{c}{\textbf{CLAQUA}}\\
     \midrule
     BM25 & \textbf{0.9600} & \textbf{0.0992} & \textbf{0.9920} & \textbf{0.9683}\\
     Doc2Query + BM25 \textcolor{red}{$\downarrow$} & 0.9395 & 0.0990 & 0.9901 & 0.9523 \\
     \bottomrule
    \end{tabular}
    \end{adjustbox}   
    \label{tab:ranking_task_performance}
\end{table}
Table \ref{tbl:clarification_need_predication} shows the results of two top-performing models (\ie BERT and RandomForest) for the clarification need prediction task ($T_1$) from traditional ML and language models. A key observation is that the prediction of clarification need should be selectively made in a classification or regression setup. In particular, BERT, a language model that well classifies the classification need on ClariQ and CLAQUA datasets, does not consistently outperform a classic approach, RandomForest, in addressing a regression-wise task (as per the results on MIMICS and MIMICS-Duo). Next, for the second sub-task, ask clarification questions, which can be addressed via generation or ranking. However, clarification question generation requires a detailed context description and associated information. The existing approaches (e.g., Seq2Seq models) could be either naive in solely taking the query as input for CQ generation or difficult to generalise to many datasets while using specific information. 
Therefore, in this study, we compare the ranking performance when applying some commonly used ranking baselines (\ie BM25 and BM25 with query expanded via the Doc2Query technique~\cite{nogueira2019document}) on every dataset. Table \ref{tab:ranking_task_performance} presents the experimental results of these two approaches on every dataset. Note that, we ignore the experimental results on ClariT, MIMICS, MIMICS-DUO and AmazonCQ since they are different from other datasets in having queries with multiple relevant clarification questions. For the results, we observe that the query expansion via Doc2Query can be effective for most of the conversational search datasets, due to their shorter queries. However, when query expansion is applied to a Conv.~QA dataset, it is not promising for an improved performance. Another observation is that the Qulac, ClariQ and ClariQ-FKw datasets have similar clarification questions in their dataset as per Figure~\ref{fig:tsne_convsearch_datasets} and Doc2Query-based query expansion has limited improvement to BM25 on these datasets. However, for another two corpus, TavakoliCQ and MANtIS, with distinct clarification questions, a bigger improvement margin can be observed. This also indicates the usefulness of our introduced visualisation-based strategy for dataset selection. 



Next, for the third task, it is crucial to determine user satisfaction with clarification questions (CQs), as it provides insight into how well the CQs are serving their intended purpose. However, obtaining the necessary data for evaluating user satisfaction can be challenging. In the literature, only two datasets (\ie MIMICS and MIMICS-Duo) include information for this task. In Table \ref{tbl:user_satisfaction_prediction}, we present the corresponding results. A similar observation to the clarification need prediction task is that the language model can assist an ACQ technique in effectively evaluating user satisfaction. However, due to the limited number of applicable datasets, this observation might not be consistent in a different context. This also aligns with the current status of the ACQ research task while evaluating the newly proposed ACQ techniques. 


Overall speaking, with the presented experimental results, we indicate the inconsistent performance of models while evaluated on different datasets. In particular, we also discuss the limited numbers of useful datasets while evaluating ACQ techniques (e.g., the models' performance on user satisfaction prediction).

\begin{table}[]
    \centering
    \caption{User satisfaction prediction with CQs performance of running best representative methods from traditional ML and language models (MultinomialNB and distilBERT) on datasets. \textcolor{green}{$\uparrow$} is added to distilBERT to indicate a consistent performance improvement on all evaluation metrics. (The results of all methods are added on Table \ref{tbl:user_satisfaction_results} in Appendix \ref{sec:exp_satisfaction_appendix}).}
    \begin{adjustbox}{width=0.8\columnwidth}
    \begin{tabular}{l|ccc}
    \toprule
    \midrule
    \textbf{Model} &  \textbf{Precision} & \textbf{Recall} & \textbf{F1}\\
    \midrule
    & \multicolumn{3}{c}{\textbf{MIMICS}}\\
    \midrule
    MultinomialNB & 0.8255 & 0.7842 & 0.7758 \\
    distilBERT \textcolor{green}{$\uparrow$} & \textbf{0.9453} & \textbf{0.9397} & \textbf{0.939} \\
     \midrule
     & \multicolumn{3}{c}{\textbf{MIMICS-Duo}}\\
     \midrule
     MultinomialNB & \textbf{0.4407} & 0.2787 & 0.2336 \\
     distilBERT & 0.2766 & \textbf{0.2803} & \textbf{0.2777} \\
     \bottomrule
    \end{tabular}
    \end{adjustbox}    
    \label{tbl:user_satisfaction_prediction}
\end{table}

\section{Discussion and Future Challenges}
\label{sec:discussion_challanges}
From the exploration of datasets as well as the experimental results on them, in this section, we highlight the concluding remarks on the current status of the ACQ research task, mainly from the dataset point of view. In addition, we discuss the promising directions based on the main findings listed below.

\textbf{Findings.} (1) \textbf{Missing Standard Benchmark.} Existing datasets are underdeveloped, and difficult to constitute a standard benchmark while introducing novel ACQ techniques. As a consequence, it is challenging to effectively and accurately compare the proposed techniques and capture the true state-of-the-art.
(2) \textbf{Few User-System Interactions Recorded for Evaluation.} In the literature, only the MIMICS dataset was collected by using a clarification pane that simulates such interactions. This makes it challenging to evaluate models in a near-realistic scenario and to estimate how well they could perform in a real-world setting.
(3) \textbf{Inconsistent Dataset Collection and Formatting.} Many included datasets in this paper are frequently presented in distinct structures and can only be applied with a tailored setup. This is a problem while developing techniques and evaluating them on multiple datasets. (4) \textbf{Inconsistent Model Evaluation.} Many newly introduced models apply customised evaluation strategies even while using an identical dataset for addressing a specific asking clarification task. This lead to difficulties in model performance comparison.

\paragraph{Future Research Directions.} (1) \textbf{Benchmark Development.} For the development of an ACQs technique, it is important that the models are compared to a common-accepted benchmark to make the corresponding conclusions. However, according to the above findings, currently, it is still unavailable. Therefore, benchmark development is the first key future direction. (2) \textbf{ACQ Evaluation Framework.} Aside from the benchmark development, it is also essential for a proper evaluation of newly introduced techniques. In particular, due to the human-machine interaction nature of the ACQ techniques, it is valuable for evaluation metrics to take user satisfaction information into account. In addition, the introduction of a corresponding evaluation framework can assist the development of ACQ techniques with systematic evaluations. (3) \textit{\textbf{Large-Scale Human-to-Machine Dataset.}} Existing datasets have many limitations that increase the difficulty of developing large-scale models for generating or ranking clarification questions. It remains challenging to collect and build large amounts of data. In the near future, researchers should optimize the process of ACQs based on the current retrieval technologies (see \cite{trippas2018informing} for a description of collecting such datasets). 
(4) \textit{\textbf{Multi-Modal ACQs Dataset.}} Recently multi-modal conversational information seeking has received attention in conversational systems \cite{deldjoo2021towards}. Amazon Alexa\footnote{\url{https://www.amazon.science/alexa-prize/taskbot-challenge}} organised the first conversational system challenge to incorporate multi-modal (voice and vision) customer experience. However, there is a lack of existing datasets containing multi-modal information for ACQs.

\section*{Limitations}
\label{sec:limitations}
In this section, we outline the key limitations of our research. Our findings on the ACQ models are not as advanced as the current state-of-the-art, but they serve as a benchmark for others to compare with when using similar datasets. Additionally, to conduct more extensive experiments on larger datasets and more advanced models, we require additional computational resources. Specifically, generating clarification questions is a demanding task as it requires the use of powerful language models.
\section*{Acknowledgments}
\label{sec:acknowledgments}
This research is supported by the Engineering and Physical Sciences Research Council [EP/S021566/1] and the EPSRC Fellowship titled ``Task Based Information Retrieval'' [EP/P024289/1].


\bibliography{anthology,custom}
\bibliographystyle{acl_natbib}

\appendix
\newpage
\section{Datasets Details}
\label{sec:datasets_details}

\subsubsection{ClariT}
\label{sec:clarit}

The ClariT dataset \cite{feng2023towards} was released in 2023 by researchers from the University College London. ClariT is the first dataset for asking clarification questions in task-oriented conversational information seeking. They built ClariT based on an existing dataset ShARC\footnote{\url{https://sharc-data.github.io}}, which clarifies users' information needs in task-oriented dialogues. They extended dialogues in ShARC with user profiles to ask clarification questions considering personalized information. To ask clarification questions efficiently, they also removed unnecessary clarification questions in the original dialogues. The collected dataset consists of over $108k$ multi-turn conversations including clarification questions, user profiles, and corresponding task knowledge in general domains.

\subsubsection{Qulac}
\label{sec:qulac}

The Qulac (\textbf{Qu}estions for
\textbf{la}ck of \textbf{c}arity) \cite{AliannejadiSigir19} dataset is a joint effort by researchers from the Università della Svizzera Italiana and the University of Massachusetts Amherst. Qulac is the first dataset as well as an offline evaluation framework for studying clarification questions in open-domain information-seeking conversational search systems. To acquire the clarification questions, they proposed a four-step strategy: (1) they defined the topics and their facets borrowed from TREC Web Track\footnote{\url{https://trec.nist.gov/data/webmain.html}}; (2) they collected several candidates clarification questions for each query through crowdsourcing in which they asked human annotators to generate questions for a given query according to the results showed using a commercial search engine; (3) they assessed the relevance of the questions to each facet and collected new questions for those facets that require more specific questions; (4) finally, they collected the answers for every query-facet-question triplet. The collected dataset consists of over $10,277$ single-turn conversations including clarification questions and their answers on multi-faceted and ambiguous queries for $198$ topics with $762$ facets.

\subsubsection{ClariQ}
\label{sec:clariq}

The ClariQ dataset \cite{aliannejadi2020convai3,aliannejadi2021building} was released in 2020 by researchers from the University of Amsterdam, Microsoft, Google, University of Glasgow, and MIPT.
The ClariQ dataset was collected as part of the ConvAI3\footnote{\url{http://convai.io}} challenge which was co-organized with the \texttt{SCAI}\footnote{\url{https://scai-workshop.github.io/2020/}} workshop. 
The ClariQ dataset is an extended version of Qulac, \ie new topics, questions, and answers have been added in the training set using crowdsourcing. Like Qulac, ClariQ consists of single-turn conversations (initial\_request, followed by clarification questions and answers). Moreover, it comes with synthetic multi-turn conversations (up to three turns). ClariQ features approximately $18K$ single-turn conversations, as well as $1.8$ million multi-turn conversations. 

\subsubsection{TavakoliCQ}
\label{sec:tavakoli2021onStackExchange}
Recently \citeauthor{tavakoli2021analyzing}~\cite{tavakoli2021analyzing,tavakoli2020generating}, from RMIT University and the University of Massachusetts Amherst, explore the ACQs to provide insightful analysis into how they are used to disambiguate the user ambiguous request and information needs. To this purpose, they extracted a set of clarification questions from posts on the \se question answering community \cite{tavakoli2020generating}. They investigate three sites with the highest number of posts from three different categories covering a period from July 2009 to September 2019. Therefore, the created dataset includes three domains, \ie business domain with $13,187$ posts, culture with $107,266$ posts, and life/arts with $55,959$ posts. To identify the potential clarification questions, they collected the comments of each post that contain at least one sentence with a question mark, excluding questions submitted by the author of the post and questions that appeared in quotation marks. Their finding indicates that the most useful clarification questions have similar patterns, regardless of the domain.

\subsubsection{MIMICS}
MIMICS (stands for the \textbf{MI}crosoft's \textbf{M}ixed-\textbf{I}nitiative \textbf{C}onversation \textbf{S}earch Data) \cite{zamani2020mimics}. This is a large-scale dataset for search clarification which is introduced in 2020 by researchers from Microsoft. Recently, Microsoft Bing added a clarification pane to its results page to clarify faceted and ambiguous queries.\footnote{However, this feature is not yet available for
some international markets.} Each clarification pane includes a clarification question and up to five candidate answers. They used internal algorithms and machine learning models based on users' history with the search engine and content analysis to generate clarification questions and candidate answers. The final MIMICS dataset contains three datasets: (1) MIMICS-Click includes $414,362$ unique queries, each related to exactly one clarification pane, and the corresponding aggregated user interaction clicks; (2) MIMICS-ClickExplore contains the aggregated user interaction signals for over $64,007$ unique queries, each with multiple clarification panes, \ie $168,921$ query-clarification pairs; (3) MIMICS-Manual includes over $2$k unique real search queries and $2.8$k query-clarification pairs. Each query-clarification pair in this dataset has been manually labeled by at least three trained annotators and the majority voting has been used to aggregate annotations. It also contains graded quality labels for each clarification question, the candidate answer set, and the landing result page for each candidate answer. 

\subsubsection{MANtIS}
\label{sec:mantis}
The MANtIS (short for \textbf{M}ulti-dom\textbf{A}i\textbf{N} \textbf{I}nformation \textbf{S}eeking dialogues) dataset \cite{penha2019introducing} is a large-scale dataset containing multi-domain and grounded information-seeking dialogues introduced by researchers from TU Delft. They built the MANtIS dataset using extraction of conversations from the \se question answering community. 
This dataset includes $14$ domains on \se. Each question-answering thread of a \se site is a conversation between an information seeker and an information provider. These conversations are included if (1) it takes place between exactly two users; (2) it consists of at least 2 utterances per user; (3) it has not been marked as spam, offensive, edited, or deprecated; (4) the provider’s utterances contain at least a reference (a hyperlink), and; (5) the final utterance belongs to the seeker and contains positive feedback. The final MANtIS dataset includes $80$k conversations over $14$ domains. Then, to indicate the type of user intent, they sampled $1,365$ conversations from MANtIS and annotate their utterances according to the user intent, such as \textit{original question}, \textit{follow-up question}, \textit{potential answer}, \textit{positive feedback}, \textit{negative feedback}, etc. The final sample contains $6,701$ user intent labels.

\subsubsection{ClariQ-FKw}
\label{sec:ClariQFKw}
The ClariQ-FKw (FKw stands for Facet Keywords) \cite{sekulic2021towards} was proposed by researchers from the University of Amsterdam and the Università della Svizzera Italiana in 2021. Their main objective was to use text generation-based large-scale language models to generate clarification questions for ambiguous queries and their facets, where by facets they mean keywords that disambiguate the query. The dataset includes queries, facets, and clarification questions, which form triplets construed on top of the ClariQ \cite{aliannejadi2020convai3} dataset. To this end, they perform a simple data filtering to convert ClariQ data samples to the appropriate triplets and derive the facets from topic descriptions. 
The final ClariQ-FKw contains $2,181$ triplets. 

\subsubsection{MSDialog}
\label{sec:msdialogue}
The MSDialog \cite{qu2018analyzing} proposed by researchers from the University of Massachusetts Amherst, RMIT University, Rutgers University, and Alibaba Group, is used to analyse information-seeking conversations by user intent distribution, co-occurrence, and flow patterns in conversational search systems. The MSDialog dataset is constructed based on the question-answering interactions between information seekers and providers on the online forum for Microsoft products. Thus, to create the MSDialog dataset, they first crawled over $35$k multi-turn QA threads (\ie dialogues) containing $300$k utterances from the Microsoft Community\footnote{\url{https://answers.microsoft.com/}} -- a forum that provides technical support for Microsoft products -- and then annotated the user intent types on an utterance level based on crowdsourcing using Amazon Mechanical Turk (MTurk)\footnote{\url{https://www.mturk.com/}}. To provide a high-quality and consistent dataset, they selected about $2.4$k dialogues based on four criteria, conversations 1) with 3 to 10 turns; 2) with 2 to 4 participants; 3) with at least one correct answer selected by the community, and; 4) that fall into one of the following categories: Windows, Office, Bing, and Skype, which are the major categories of Microsoft products. The final annotated dataset contains $2,199$ multi-turn dialogues with $10,020$ utterances.

\subsubsection{MIMICS-Duo}
\label{sec:mimics-duo}
The MIMICS-Duo \cite{tavakoli2022mimics} dataset is proposed by researchers at RMIT University, the University of Melbourne, and the University of Massachusetts Amherst. It provides the online and offline evaluation of clarification selection and generation approaches. It is constructed based on the queries in MIMICS‐ClickExplore \cite{zamani2020mimics}, a sub-dataset of MIMICS \cite{zamani2020mimics} that consists of online signals, such as user engagement based on click‐through rate. The MIMICS‐Duo contains over $300$ search queries and $1,034$ query‐clarification pairs.

\subsubsection{ClarQ}
\label{sec:clarq}
The ClarQ dataset \cite{kumar2020clarq} was created in 2020 by Carnegie Mellon University. The ClarQ is designed for large-scale clarification question generation models. To do this, the ClarQ dataset is built with a bootstrapping framework based on self supervision approaches on top of the post-comment tuples extracted from \se\footnote{\url{https://stackexchange.com/}} question answering community. To construct the ClarQ, they first extracted the posts and their comments from $173$ domains. Then, they filtered unanswered posts and only considered comments to posts with at least one final answer as a potential candidate for a clarification question. The ClarQ dataset consists of about $2$ million 
post-question tuples across $173$ domains.

\subsubsection{RaoCQ}
\label{sec:rao2018onStackExchange}
\citeauthor{rao2018learning} [\citeyear{rao2018learning}] from the University of Maryland study the problem of ranking clarification questions and propose an ACQs dataset on top of StackExchange. 
To create this dataset, they use a dump of \se and create a number of post-question-answer triplets, where the post is the initial unedited request, the question is the first comment containing a question (\ie indicated by a question mark), and the answer is either the edits made to the post after the question (\ie the edit closest in time following the question) or the author's answer of the post to the question in the comment section. The final dataset includes a total of $77,097$ triples across three domains \textit{askubuntu}, \textit{unix}, and \textit{superuser}. 

\subsubsection{AmazonCQ}
\label{sec:rao2019onAmazon}
\citeauthor{rao2019answer} [\citeyear{rao2019answer}] from Microsoft and the University of Maryland, released a dataset for generating clarification questions. The dataset contains a context that is a combination of product title and description from the Amazon website,
a question that is a clarification question asked to the product about some missing information in the context, and the answer that is the seller's (or other users') reply to the question. To construct this dataset, they combined the Amazon Question Answering dataset created by \cite{mcauley2016addressing} and the Amazon Review dataset proposed by \cite{mcauley2015image}. The final dataset consists of $15,859$ contexts (\ie product description) with $3$ to $10$ clarification questions, on average $7$, per context.

\subsubsection{CLAQUA}
\label{sec:claqua}
The CLAQUA dataset \cite{xu2019asking} was created by researchers from of Peking University, the University of Science and Technology of China, and Microsoft Research Asia in 2019. They propose the CLAQUA dataset to provide a supervised resources for training, evaluation and creating powerful models for clarification-related text understanding and generation in knowledge-based question answering (KBQA) systems. The CLAQUA dataset is constructed in three steps, (1) sub-graph extraction, (2) ambiguous question annotation, and (3) clarification question annotation. In the first step, they extract ambiguous sub-graphs from an open-domain knowledge base, like FreeBase. They focus on shared-name ambiguity where two entities have the same name and there is a lack of necessary distinguishing information. Then, in the second step, they provide a table listing the shared entity names, their types, and their descriptions. Based on this table, annotators need to write ambiguous questions.
Finally, in the third step, based on entities and the annotated ambiguous question, annotators are required to summarize distinguishing information and write a multi-choice clarification question including a spacial character that separate entity and pattern information.
They provided these steps for single- and multi-turn conversations. The final CLAQUA dataset contains $17,163$ and $22,213$ single-turn and multi-turn conversations, respectively.


\section{Experiments on Model Performance}
\label{sec:exp_model_performance}

\subsection{Clarification Need Prediction}
\label{sec:exp_need_prediction_appendix}
The clarification need prediction is a major task in search clarification to decide whether to ask clarification questions. Between the discussed CQ datasets only ClariQ \cite{aliannejadi2020convai3,aliannejadi2021building}, MIMICS \cite{zamani2020mimics}, MIMICS-Duo \cite{tavakoli2022mimics}, and CLAQUA \cite{xu2019asking} provide the necessary information for the clarification need prediction task. The ClariQ and CLAQUA datasets model the clarification need prediction task as a classification problem. They both present the initial user request with a classification label that indicates the level of clarification required. In contrast to the ClariQ and CLAQUA datasets, the task in the MIMICS and MIMICS-Dou datasets is modelled as a regression task for predicting user engagement. Specifically, these datasets aim to predict the degree to which users find the clarification process useful and enjoy interacting with it. Based on this prediction, the system can make a decision on whether or not to request clarification. We subsequently evaluated the prediction task for clarification needs using a variety of traditional machine learning models and language models. The traditional machine learning models employed as baselines include Random Forest \cite{breiman2001random}, Decision Tree \cite{loh2011classification}, Multinomial Naive Bayes (MultinomialNB) \cite{manning2008introduction}, Support Vector Machines (SVM) \cite{cortes1995support}, and Linear Regression \cite{yan2009linear}. The language model baselines utilized include BART \cite{lewis2019bart}, XLNet \cite{yang2019xlnet}, XLM \cite{lample2019cross}, Albert \cite{lan2019albert}, distilBERT \cite{sanh2019distilbert}, and BERT \cite{devlin2018bert}. These models were applied to both classification and regression tasks. The input to traditional ML models is a matrix of TF-IDF features extracted from the raw input text. We use Scikit-learn\footnote{\url{https://scikit-learn.org/}} \cite{scikit-learn}, HuggingFace\footnote{\url{https://huggingface.co/}} \cite{wolf2019huggingface}, and TensorFlow \cite{abadi2016tensorflow} for the implementation of the aforementioned models.

\subsection{Question Relevance Ranking Baselines}
To address the second task, namely asking clarification questions, many studies have explored either generation or ranking strategies. However, as we argued in Section~\ref{sec:performance}, the generation techniques require rich information for satisfactory performance and they are difficult to be applied to many datasets if some specific information is required. Therefore, we consider the ranking task for summarsing the model performance on the asking clarification 
question task and present the results of BM25 and Doc2Query + BM25. Note that, the BM25-based techniques are considered with their competitive performance in addressing the clarification question ranking task~\cite{aliannejadi2021building}. We also compare some additional ranking techniques, such as the PL2~\cite{amati2002probabilistic}, DPH~\cite{amati2008fub} and another recent dense retriever (\ie ColBERT~\cite{khattab2020colbert}). However, the inclusion of such approaches is not useful while comparing the use of different datasets. Therefore, we only present the results of the above two approaches in Table~\ref{tab:ranking_task_performance}. As for the implementation, we leverage PyTerrier\footnote{\url{https://github.com/terrier-org/pyterrier}} \cite{pyterrier2020ictir}, a recently developed Python framework for conducting information retrieval experiments. 

\subsection{User Satisfaction with CQs}
\label{sec:exp_satisfaction_appendix}
In this experiment, we explored the task of determining user satisfaction with CQs by utilizing a variety of models from both traditional machine learning and language models on the ACQs datasets. To conduct this experiment, we employed the same models that we previously used for the Clarification Need Prediction task. By using the same models for both tasks, we aim to examine how well these models perform in predicting user satisfaction with CQs and how their performance compares to their performance in predicting the need for clarification. This will allow us to understand the strengths and limitations of these models in predicting user satisfaction and make informed decisions on which models to use in future applications. Only two datasets (\ie MIMICS \cite{zamani2020mimics} and MIMICS-Duo \cite{tavakoli2022mimics}) out of $12$ datasets provide the user satisfaction information. In both MIMICS and MIMICS-Dou, each clarification question is given a label to indicate how a user is satisfied with the clarification question. For MIMICS the labels are Good, Fair, or Bad. A good clarifying question is accurate, fluent, and grammatically correct. A fair clarifying question may not meet all of these criteria but is still acceptable. Otherwise, it is considered bad. While in MIMICS-Dou, users' satisfaction with clarification questions is assessed on a $5$‐level scale that is Very Bad, Bad, Fair, Good, and Very Good. Thus, we formulate user satisfaction with CQs task as a supervised classification in our experiments.

\begin{table*}
    \centering
    \caption{The performance of all methods on clarification need prediction on MIMICS and MIMICS-Duo. The best models are in \textbf{bold}.}
    \begin{tabular}{lccclccc}
    \toprule
    \multirow{2}{*}{\textbf{Model}} & \multicolumn{3}{c}{\textbf{MIMICS}} && \multicolumn{3}{c}{\textbf{MIMICS-Duo}} \\
    \cline{2-4}\cline{6-8}
    & Precision & Recall & F1 && Precision & Recall & F1 \\
     \midrule
     RandomForest & 0.3540 & \textbf{0.3806} & \textbf{0.3717} && 0.2860 & 0.5000 & 0.3638 \\
     DecisionTree & 0.2125 & 0.2520 & 0.2028 && 0.5329 & 0.5095 & 0.4305\\
     SVM & 0.2858 & 0.3024 & 0.2772 && 0.5281 & 0.5088 & 0.4333\\
     MultinomialNB & 0.2924 & 0.3186 & 0.2876 && 0.5185 & 0.5178 & 0.5166\\
     LogisticRegression & 0.2749 & 0.2878 & 0.2816 && \textbf{0.7862} & 0.5010 & 0.3660\\
     \hline
     BART & 0.5083 & 0.3344 & 0.3657 && 0.5869 & 0.5503 & 0.5194 \\
     XLNet & 0.1385 & 0.2500 & 0.1782 && 0.286 & 0.5 & 0.3638  \\
     XLM & 0.0119 & 0.2500 & 0.0227 && 0.286 & 0.5 & 0.3638  \\
     Albert & 0.2920 & 0.2877 & 0.2855 && 0.286 & 0.5 & 0.3638\\
     distilBERT & 0.3391 & 0.3305 & 0.3322 && 0.5941 & 0.594 & 0.5941 \\
     BERT & \textbf{0.3804} & 0.3249 & 0.3344 && 0.6349 & \textbf{0.625} & \textbf{0.6255} \\
     \midrule
     \multirow{1}{*}{\textbf{}} & \multicolumn{3}{c}{\textbf{MIMICS}} && \multicolumn{3}{c}{\textbf{MIMICS-Duo}} \\
     \cline{2-4}\cline{6-8}
    & MAE & MSE & R$^2$ && MAE & MSE & R$^2$ \\
    \midrule
     RandomForest & 2.4404 & 7.969 & -0.0012 && 2.8502 & 11.206 & \textit{-0.0079}\\
     DecisionTree & 2.6374 & 10.0143 & -0.2581 && 3.052 & 14.2306 & -0.2799\\
     SVR & 2.4447 & 8.1852 & -0.0283 && 2.7801 & 14.6398 & -0.3167\\
     MultinomialNB & 3.3364 & 16.7424 & -1.1034 && 2.7971 & 18.942 & -0.7037\\
     LogisticRegression & 3.4084 & 17.9488 & -1.2549 & & 2.7971 & 18.942 & -0.7037\\
     \hline
     BART & 2.3903 & 8.5296 & -0.0716 && \textbf{2.7233} & \textbf{10.3239} & \textbf{0.0714}\\
     XLNet & 2.4582 & 8.1836 & -0.0281 && 2.7971 & 18.942 & -0.7037\\
     XLM & 2.6214 & 9.9151 & -0.2456 && 2.7971 & 18.942 & -0.7037\\
     Albert & 2.4339 & 8.0300 & -0.0088 && 2.7971 & 18.942 & -0.7037\\
     distilBERT & \textbf{2.3325} & \textbf{7.8685} & \textbf{0.0115} && 2.7744 & 11.0613 & 0.0051\\
     BERT & 2.4562 & 8.1277 & -0.0211 && 2.8801 & 11.2268 & -0.0098\\
     \bottomrule
    \end{tabular}
    \label{tbl:clarificationneed_prediction_results}
\end{table*}

\begin{table*}
    \centering
    \caption{The performance of all methods on user satisfaction prediction with CQs on MIMICS and MIMICS-Duo. The best models are in \textbf{bold}.}
    \begin{tabular}{lccclccc}
    \toprule
    \multirow{2}{*}{\textbf{Model}} & \multicolumn{3}{c}{\textbf{MIMICS}} && \multicolumn{3}{c}{\textbf{MIMICS-Duo}} \\
    \cline{2-4}\cline{6-8}
    & Precision & Recall & F1 && Precision & Recall & F1 \\
     \midrule
     RandomForest & 0.7522 & 0.5172 & 0.3686 && 0.1256 & 0.25 & 0.1672 \\
     DecisionTree & 0.5648 & 0.5168 & 0.4050 && 0.2218 & 0.2311 & 0.2163\\
     SVM & 0.736 & 0.5947 & 0.5212 && 0.2379 & 0.2498 & 0.2157\\
     MultinomialNB & 0.8255 & 0.7842 & 0.7758 && \textbf{0.4407} & 0.2787 & 0.2336\\
     LogisticRegression & 0.7522 & 0.5172 & 0.3686 && 0.3762 & 0.2542 & 0.1761\\
     \midrule
     BART & 0.9385 & 0.931 & 0.9302 && 0.1256 & 0.25 & 0.1672\\
     XLNet & 0.9219 & 0.9217 & 0.9217 && 0.1256 & 0.25 & 0.1672\\
     XLM & 0.9348 & 0.9309 & 0.9303 && 0.1256 & 0.25 & 0.1672\\
     Albert & 0.9385 & 0.931 & 0.9302 &&  0.1256 & 0.25 & 0.1672\\
     distilBERT & \textbf{0.9453} & \textbf{0.9397} & \textbf{0.939} && 0.2766 & \textbf{0.2803} & \textbf{0.2777} \\
     BERT & 0.9385 & 0.931 & 0.9302 && 0.2851 & 0.264 & 0.2056\\
     \bottomrule
    \end{tabular}
    \label{tbl:user_satisfaction_results}
\end{table*}

\end{document}